# High-sensitivity mapping of magnetic induction fields with nanometer-scale resolution: comparison of off-axis electron holography and pixelated differential phase contrast

Victor Boureau[1,2], Michal Staňo[3,4], Jean-Luc Rouvière[5], Jean-Christophe Toussaint[3], Olivier Fruchart[6] and David Cooper[1]

1: Univ. Grenoble Alpes, CEA, LETI, 38000 Grenoble, France.
2: Interdisciplinary Center for Electron Microscopy (CIME), EPFL, CH-1015 Lausanne, Switzerland.
3: Univ. Grenoble Alpes, CNRS, Institut Néel, 38000 Grenoble, France.
4: CEITEC BUT, Brno University of Technology, 612 00 Brno, Czech Republic.
5: Univ. Grenoble Alpes, CEA, IRIG-MEM, 38000 Grenoble, France.
6: Univ. Grenoble Alpes, CNRS, CEA, Spintec, 38000 Grenoble, France.

**Abstract**

We compare two transmission electron microscopy (TEM) based techniques that can provide highly spatially resolved quantitative measurements of magnetic induction fields at high sensitivity. To this end, the magnetic induction of a ferromagnetic NiFe nanowire has been measured and compared to micromagnetic modelling. State-of-the-art off-axis electron holography has been performed using the averaging of large series of holograms to improve the sensitivity of the measurements. These results are then compared to those obtained from pixelated differential phase contrast (DPC), a technique that belongs to pixelated (or 4D) scanning transmission electron microscopy (STEM) experiments. This emerging technique uses a pixelated detector to image the local diffraction patterns as the beam is scanned over the sample. For each diffraction pattern, the deflection of the beam is measured and converted into magnetic induction, while scanning the beam allows a map to be generated. Aberration corrected Lorentz (field-free) configurations of the TEM and STEM were used for an improved spatial resolution. We show that the pixelated STEM approach, even when performed using an old generation of charge-coupled device camera, provides better sensitivity at the expense of spatial resolution. A more general comparison of the two quantitative techniques is given.

**Introduction**

A number of three-dimensional magnetization objects can have length scales below 10 nm, such as vortex cores [1], skyrmions [2] and Bloch points [3]. These are textbook cases for the fundamental study of nanomagnetics and magnetization dynamics, and also building blocks for novel spintronic devices, such as spin-torque oscillators [4], data storage, logic and unconventional electronics [5], up to three-dimensional memories [6]. As such, one needs characterization techniques that can probe these magnetization textures quantitatively with a spatial resolution of a few nanometers, and with high sensitivity to provide a sufficient signal-to-noise ratio of the weak magnetic signal generated by the low amount of material. Magnetic force microscopy (MFM) is a widespread technique used to probe the magnetic stray field at the surface of a sample, but the quantification of the measurement remains a challenge [7]. On the other hand, two rapidly-developing techniques are progressively meeting these criteria. First, X-ray dichroism ptychography [8], which probes the magnetic induction component that is parallel to the beam. Secondly, transmission electron microscopy (TEM) based techniques that use a Lorentz optical configuration [9], which probe the magnetic induction component that is perpendicular to the beam. In this paper, we discuss two of the main techniques from this latter category capable of delivering quantitative maps of magnetic induction fields at the nanoscale. The recently developed pixelated differential phase contrast (DPC) technique [10–12] is compared to the well-established off-axis electron holography technique [13,14].

Off-axis electron holography is a TEM method that has been used for quantitative mapping of magnetic vector potentials at the nanoscale since the 1980s, utilizing the Ehrenberg-Siday-Aharonov-Bohm effect [15,16]. A schematic of the experimental setup for off-axis electron holography is shown in Figure 1(a). An electron biprism is used to interfere two coherent electron beams: the object wave that has passed through the sample, with the reference wave that has passed through a region free of electromagnetic potential. From the resulting interference pattern, known as a hologram, the amplitude and phase of the electron beam can be mapped using a simple Fourier reconstruction procedure, from which a magnetic induction map can be obtained. In this manuscript, we will refer to this technique simply as "electron holography", although there are many different variations of lesser-known electron holography-based techniques [17], notably in-line electron holography which has been proven for magnetic measurements [18,19].

Pixelated (also called 4D) scanning transmission electron microscopy (STEM) techniques have become widespread in recent years due to the availability of hybrid pixel array detectors [20,21], making possible the rapid recording of diffraction patterns synchronized with the scanning of the STEM probe on the sample [22]. A schematic of the experimental setup is shown in Figure 1(b). While interacting with the sample, the convergent electron beam is deflected by the gradient of the magnetic vector potential [9]. Pixelated DPC measures the shift of the transmitted beam in each diffraction pattern. It is related to the deflection angle of the beam and converted into a local magnetic induction field. In this way, pixelated DPC can be seen as an evolution of the well-known DPC technique that uses an annular segmented detector to record the deflection of the transmitted beam [23–26]. However, the deflection of the beam is usually not a rigid shift of the intensity distribution of the diffraction pattern and as such quantitative measurement by DPC is complicated [27,28].

An isolated ferromagnetic $Ni_{0.86}Fe_{0.14}$ nanowire of 56 nm diameter has been used to benchmark these two techniques. Using this structure, in a single field of view, we may measure the magnetization in the material along the wire, the non-uniform magnetization textures at its apex, and stray fields around the wire.

**Specimen fabrication**

Cylindrical NiFe nanowires were fabricated by electroplating in an aqueous solution inside a nanoporous alumina template. The bottom part of the pores with 50 nm nominal diameter was covered by a thin Au layer that served as a working electrode. The deposition was carried out at a potential of -1.6 V compared to the saturated calomel reference electrode, for 300 s using a Pt counter electrode. The solution (pH 3) consisted of 6 g/l $FeSO_4 \cdot 7H_2O$, 26 g/l $NiSO_4 \cdot 6H2O$, 25 g/l $H_3BO_3$ and 3 g/l $C_6H_8O_6$ (ascorbic acid) [29]. After the deposition, the template was dissolved in 1 M NaOH to free the nanowires. After rinsing with demineralized water and isopropyl alcohol, the wires were dispersed on a lacey carbon grid for the TEM observations.

**Electron holography measurements**

The off-axis electron holography experiments were performed using a FEI Titan Ultimate microscope, operated at 200 kV. When performing magnetic mapping in a TEM, a field-free environment is required such that the field from the magnetic lenses does not influence the distribution of magnetization inside the sample. This was achieved by switching off the conventional objective lens and using a Lorentz lens configuration. The additional use of an image aberration corrector provided a spatial resolution of better than 1 nm by correcting optical aberrations up to the third order, compared to that of standard Lorentz imaging which is of the order of 3 nm [30]. To provide improved phase sensitivity, series of 100 electron holograms, each with an exposure time of 8 s and a fringe spacing of 1.1 nm, were acquired using a complementary metal-oxide-semiconductor (CMOS) camera (Gatan OneView) located right below the viewing screen. A series of reference holograms was used to correct the phase distortions of the optical system. The aligned and averaged phase and amplitude images were then reconstructed from the hologram series using the Holoview software [31].

The phase $\phi$ of an electron travelling along the $z$ direction will be modified as it passes through both an electric potential $V$ and a magnetic vector potential $\mathbf{A}$ [14,15],

$$\phi(x,y) = \frac{e}{\hbar v}\int_{-\infty}^{+\infty} V(\mathbf{r})dz - \frac{e}{\hbar}\int_{-\infty}^{+\infty} A_z(\mathbf{r})dz \quad (1)$$

where $\mathbf{r} = x, y, z$ is the three-dimensional position vector, $A_z$ is the projection of magnetic vector potential in the $z$ direction, $e$ is the elementary charge, $v$ is the relativistic electron velocity and $\hbar$ is the reduced Planck's constant. Thus, the phase images measured by off-axis electron holography contain two contributions, the former related to the electric potential and the latter related to the magnetic vector potential. The separation of these two contributions is required for proper analysis. Here, we achieve it using the "flipping method", by performing the holography experiment with two orientations of the specimen, flipping the specimen by 180° to provide a second phase image [32]. As the magnetic vector potential is then reversed, the difference between the two phase images is two times the magnetic phase contribution, while the electric phase is half of the sum of the two phase images. Then the projection of the magnetic induction field $\mathbf{B}$ and electric field $\mathbf{E}$ can be calculated from the gradient of the corresponding phase images using the following equations. The magnetic induction in Coulomb gauge is,

$$\mathbf{B}(\mathbf{r}) = \nabla \times \mathbf{A}(\mathbf{r}) \quad (2)$$

and the conservative electric field is,

$$\mathbf{E}(\mathbf{r}) = -\nabla V(\mathbf{r}) \quad (3)$$

where $\nabla$ is the nabla operator and $\times$ the cross product. The projection of the information measured by TEM only allows calculating the two-dimensional in-plane components of the magnetic induction $\mathbf{B}_\perp$ and of the electric field $\mathbf{E}_\perp$, that are integrated along the electron beam propagation. The subscript $\perp$ denotes the projection of the vector in the plane perpendicular to the direction of propagation of the beam.

Figures 2(a) and (b) show holography measurements of the amplitude images of the nanowire in both orientations. The phase images for each orientation containing both the magnetic and electric contributions are shown in Figures 2(c) and (d) respectively. The asymmetric phase map relative to the axis of the nanowire denotes the presence of a magnetic contribution of the specimen. Figures 2(e) and (f) show the magnetic and electric phase maps, respectively, separated by the "flipping method". The phase map shown in Figure 2(d) has been horizontally flipped in order to perform the sum and difference of both phase images. The magnetic phase map of Figure 2(e) contains information of the magnetic potential vector of both the magnetic field and the magnetization inside the NiFe material. On the other hand, the electric phase map of Figure 2(f) contains two different types of information. One is the electric potential resulting from the charging by the electron beam of the thin insulating native oxide layer at the surface of the nanowire, another one is the mean inner potential (MIP) of the material; the latter dominates at the location of the nanowire. An artifact is observed in the phase maps close to the apex of the nanowire. It results from a local loss of contrast of the hologram fringes that arise from dynamical diffraction effects, also demonstrated by the dark contrast in the amplitude map shown in Figure 2(b).

**Modelling of the magnetic phase map**

Bright-field TEM observations were used to measure the nanowire dimensions to serve as an input for the simulations. The diameter of the metallic NiFe nanowire is 56 nm, covered with a 3 nm-thick native oxide shell showing a protrusion at the sharp end of the cylinder-shaped nanowire [visible in Figure 2(a) and depicted in Figure 2(g)]. As will be demonstrated later in this section, this native oxide has no influence on the magnetic configuration. The equilibrium 3D micromagnetic configuration of the NiFe nanowire was simulated using the finite element FeeLLGood code [33,34]. Only dipolar and exchange interactions were taken into account with the following parameters: spontaneous magnetic induction $\mu_0 M_s$ = 0.9 T, exchange stiffness $A$ = 10 pJ·m$^{-1}$, wire diameter = 56 nm and length = 600 nm. The magnetic surface charges were compensated at one end of the wire as boundary conditions, to mimic a semi-infinite wire. The tetrahedron cell size was 3 nm or smaller. A simulation volume with dimensions larger than the wire segment length was chosen in order to visualize the magnetic stray field contribution and limit possible artifacts resulting from the finite size of the simulation box. The equilibrium 3D magnetic configuration, consisting of largely uniform magnetization along the wire with an end-curling at the free apex, was then post-processed to provide the projected magnetic phase map as measured by electron holography [35].

Figure 3(a) shows the simulated magnetic phase shift generated by the nanowire. A profile has been extracted from the region indicated by the dashed line, to compare in Figure 3(b) the simulation to the experimentally measured magnetic phase map. By fitting the change of magnetic phase across the nanowire using the post-processed micromagnetic model to the holography results, it is possible to determine the experimental magnetization of the NiFe alloy, of 0.9 T. It corresponds to of a Fe content of 14% [36]. Figures 3(c) and (d) show the magnetic equipotential lines from the model and the holography measurement respectively, with a smaller field of view shown in Figures 3(e) and (f). The magnetic equipotential lines depict the magnetic induction field lines and are simply displayed as the cosine of the magnetic phase, in this case amplified by a factor 30. Overall, simulated and experimental induction line maps show a good agreement. In Figure 3(d), a slight bending of the experimental magnetic stray field lines is observed in the vicinity of the bottom-left corner of the map relatively to the simulation [shown in Figure 3(c)]. It most likely results from a small measurement artifact in this region showing a very weak magnetic field, that is generated by local electrostatic charging of dirt on the electron biprism. The blue shaded region in Figures 3(e) and (f) depicts the location of the NiFe nanocylinder-shaped material, where narrow field lines resulting from the magnetization of the material can be observed running along the nanowire axis. The magnetic stray field measured in the vicinity of the magnetic nanocylinder is not altered by the protrusion at the apex of the nanowire, confirming that it has no significant influence on the magnetic configuration. The red shaded region in Figures 3(f) indicates the presence of a local artifact for the holography measurement [discussed along with Figure 2(e)], which perturbates the subsequent field lines.

**Pixelated DPC measurements**

The pixelated DPC experiments were performed using a FEI Titan Themis microscope operated at 200 kV. Observations were performed in field-free conditions, or Lorentz configuration of the STEM, where the objective lens of the microscope was switched off. This specific low magnification (LM) STEM mode was corrected for the probe aberrations up to the third order [37], so that the size of the STEM probe, which determines the spatial resolution, is essentially diffraction-limited. In these experiments, a standard charge-coupled device (CCD) camera (Gatan US1000), located after a Gatan Imaging Filter (GIF) used in non-filtering imaging mode, was used to record diffraction patterns with a 9.1 m camera length and a 300 µrad convergence semi-angle. This value of the convergence angle leads to a probe diameter of ∼ 5 nm for a diffraction-limited system, as defined by the first minimum of the Airy distribution of the probe [38]. The shift of the transmitted beam $S$ that is measured in the diffraction patterns is related to the deflection angle of the beam $\gamma$ through the camera length of the microscope $L$, written with the small angle approximation,

$$S = L.\arctan(\gamma) \approx \gamma L . \qquad (4)$$

Thus, the camera length magnifies the measurement of the beam deflection and as such a longer camera length improves the sensitivity of the measurement [39]. It is worth noticing that a compromise between the camera length and convergence semi-angle of the optical setup must be chosen, as the combination of these two parameters permits the transmitted beam to fit onto the physical size of the pixelated detector located in the diffraction plane. For these experiments, the choice of convergence semi-angle provided a sufficient spatial resolution for the measurements but limited the usable camera length to 9.1 m. 2D diffraction patterns were acquired for a 2D grating of 60 x 60 probe positions on the sample, with a step size of 9 nm to provide a similar field of view as for the holography experiments. Manageable sized 4D datasets were provided by binning the 2048 x 2048 pixels camera by eight to record the diffraction patterns as 256 x 256 pixels images, providing 0.92 Gb datasets. A dwell time of 80 ms was used for a total acquisition time per dataset of 6 min 53 s. 2D deflection maps (of 60 x 60 pixels) were obtained by calculating the shift of the transmitted beam for each diffraction pattern, relative to a reference position. The reference position can be defined as the position of the transmitted beam associated with a probe location where the sample is free of fields. The scanning system and imperfections of the descan system of the STEM produce systemic rigid shifts of the diffraction patterns, so that a reference scan (dataset acquired with the same experimental conditions, but in vacuum with no specimen) is required to infer the shift arising from the influence of the sample itself. In this study, the reference scan was directly used to obtain a 2D map of reference positions, such that no fields free location of the sample is required. Finally, the deflection maps were corrected for the rotation between the STEM imaging of the sample and the diffraction plane, by removing this orientation offset.

The shift of the beam, $\mathbf{S^{COM}}$, was measured with the center of mass (COM) of the intensity distribution of the transmitted electron beam in the diffraction plane, for each position of the STEM probe $\mathbf{R} = \mathbf{r}_\perp = x, y$. It can be shown that, considering an incident electron probe described by the wave function $\Psi$, a thin sample and under the phase object approximation (POA) [40–42],

$$\mathbf{S^{COM}}(\mathbf{R}) = \frac{1}{2\pi} |\Psi(\mathbf{r}_\perp)|^2 \star \nabla \phi(\mathbf{r}) \qquad (5)$$

where $\star$ denotes the cross-correlation and $|\Psi(\mathbf{r}_\perp)|^2 = I(\mathbf{r}_\perp)$ is the intensity distribution of the electron probe before interaction with the specimen. After substitution of the phase shift induced by the sample given by the quantum theory and shown in Equation 1, we obtain,

$$\mathbf{S^{COM}}(\mathbf{R}) = \frac{e}{h} I(\mathbf{r}_\perp) \star \left[ \frac{1}{v} \int_{-\infty}^{+\infty} \nabla V(\mathbf{r}) dz - \int_{-\infty}^{+\infty} \nabla A_z(\mathbf{r}) dz \right]. \qquad (6)$$

The use of Equations 2 and 3 leads to [43],

$$\begin{bmatrix} S_x^{COM}(\mathbf{R}) \\ S_y^{COM}(\mathbf{R}) \end{bmatrix} = -\frac{e}{h} I(\mathbf{r}_\perp) \star \left\{ \frac{1}{v} \int_{-\infty}^{+\infty} \begin{bmatrix} E_x(\mathbf{r}) \\ E_y(\mathbf{r}) \end{bmatrix} dz + \int_{-\infty}^{+\infty} \begin{bmatrix} -B_y(\mathbf{r}) \\ B_x(\mathbf{r}) \end{bmatrix} dz \right\} \qquad (7)$$

where the $x$ and $y$ subscripts denote the projection of the vector in the $x$ and $y$ directions, respectively. In case the phase gradient, i.e. $\mathbf{E}$ and $\mathbf{B}$ fields, are uniform over the spatial extension of the electron probe, the intensity distribution of the transmitted beam in the diffraction plane, $I(\mathbf{k}_\perp) = \mathrm{FT}[I(\mathbf{r}_\perp)]$ undergoes a rigid shift [27]. FT denotes the Fourier transform and $\mathbf{k}_\perp$ is the two-dimensional reciprocal space vector. Template matching (TM) algorithms are appropriate for the measurement of this rigid shift relative to a reference position, $\mathbf{S^{TM}}$ [11,44]. It can be deduced from Equation 6 or Equation 7, and expressed as a function of the gradient of potentials or fields, respectively,

$$\mathbf{S}^{TM}(\mathbf{R}) = \begin{bmatrix} S_x^{TM}(\mathbf{R}) \\ S_y^{TM}(\mathbf{R}) \end{bmatrix} = -\frac{e}{hv}\int_{-\infty}^{+\infty}\begin{bmatrix} E_x(\mathbf{r}) \\ E_y(\mathbf{r}) \end{bmatrix}dz + \frac{e}{h}\int_{-\infty}^{+\infty}\begin{bmatrix} B_y(\mathbf{r}) \\ -B_x(\mathbf{r}) \end{bmatrix}dz. \quad (8)$$

Lastly, it is interesting to show that Equation 8 which describes the rigid-shift model corresponds to the deflection of an electron travelling across an electromagnetic field via the Lorentz force $\mathbf{F}_L = -e(\mathbf{E} + \mathbf{v} \times \mathbf{B})$ in the classical mechanics formalism [9,39],

$$\boldsymbol{\gamma} = \begin{pmatrix} \gamma_x \\ \gamma_y \end{pmatrix} = -\frac{e\lambda}{hv}\int_{-\infty}^{+\infty}\begin{bmatrix} E_x(\mathbf{r}) \\ E_y(\mathbf{r}) \end{bmatrix}dz + \frac{e\lambda}{h}\int_{-\infty}^{+\infty}\begin{bmatrix} B_y(\mathbf{r}) \\ -B_x(\mathbf{r}) \end{bmatrix}dz \quad (9)$$

where $\lambda$ is the relativistic wavelength and $\boldsymbol{\gamma}$ is considered a 2D vector. Equation 9 can be directly obtained from Equation 8 as the deflection angle of the electron is given by the ratio of the momentum components $\mathbf{p}$,

$$\boldsymbol{\gamma} \approx \frac{p_\perp}{p_z} = \lambda \mathbf{k}_\perp^e \quad (10)$$

where $\mathbf{k}_\perp^e$ is the in-plane component of the wavevector of an electron that is deflected by the sample, and corresponds to the rigid shift of the beam that is measured in the diffraction plane, being a 2D vector of the reciprocal space. Note that, similarly to the electron holography experiments, only the integral of the fields along the electron beam propagation is measured. As with the electron holography experiments, the "flipping method" was adapted for pixelated DPC, using a cross-correlation based algorithm with the two virtual bright-field images to tune the alignment parameters. The application of this alignment to the maps of deflection angle allows the separation of the magnetic deflection from the electric deflection contributions (see Equations 9). To achieve this for the 2D vector maps, the horizontal component of the vectors was reversed when the map is horizontally flipped, to superpose the specimen in both images.

Figures 4(a) and (b) show the two virtual bright-field images of the nanowire that have been reconstructed by summing the intensity enclosed inside the transmitted beam (so-called virtual bright field), for both orientations of the sample used for the "flipping method". Figures 4(c-f) show the diffraction patterns recorded at four different locations of the STEM probe, as indicated in Figure 4(b). When the probe is far from the nanowire, in position 1, the transmitted beam is a disk of homogeneous intensity of radius equal to the convergence semi-angle. When the electron probe is scanned closer to the nanowire, the distribution of intensity breaks the rigid-shift model and becomes more complex, which would require dedicated modelling for a comprehensive interpretation. It demonstrates the need for using the COM instead of the TM approach for this sample. The electron probe undergoes large variations of phase gradients while interacting with the cylindrical nanowire, resulting from the MIP of the material [visible in Figure 2(f)] that is proportional to the projected thickness of the wire. For example, when the probe is on the edge of the wire, its intensity distribution shown in Figure 4(e) is largely displaced in the direction perpendicular to the wire axis [see Figure 4(i) for the orientation of the nanowire in the diffraction plane] and can be described by Equation 6. At the same time, a halo of the transmitted beam remains in the same location and results from the interaction of the tail of the electron probe with smaller phase gradients existing outside the material [42,45]. While the probe is scanned further from the nanowire in vacuum, ripples are observed in the intensity distribution of the transmitted beam and seen in Figure 4(d). They may be attributed to variations of the phase gradient along the electron beam propagation, induced by the long-range magnetic stray field generated by the magnetic wire.

Figures 4(g) and (h) show the shift of the transmitted beam inferred with a COM algorithm for each orientation of the sample, respectively. It is important to note that vectors have been corrected from an angular offset existing between the STEM image of the sample [shown in Figures 4(a) and (b)] and the diffraction that is recorded by the camera. For this purpose, an under-focused diffraction pattern is acquired, revealing the shadow image of the nanowire shown in Figure 4(i). Figures 4(j) and (k) show the deflection of the beam, where the shifts from the reference scan have been subtracted. Figure 4(l) shows the rigid shifts of the reference scan, used for obtaining Figure 4(k). At this stage, the maps contain both the magnetic and electric components of deflection. Figure 4(m) shows the magnetic component of the beam deflection and Figure 4(n) its electric component after separation by the "flipping method" that has been adapted for pixelated DPC.

## Analysis and discussion

Figure 5 compares the quantitative measurements of the fields with the simulation. Figure 5(a) and (b) show the measurements of the magnetic induction field for holography and pixelated DPC, calculated respectively from the gradient of the magnetic phase shift as described by Equations 1 and 2 and from the beam deflection as described by Equation 9. This can be compared to Figures 5(c) showing the micromagnetic simulation. Profiles extracted from the regions indicated in Figures 5(a-c) have been plotted for the holography, pixelated DPC and simulated maps of the projected magnetic induction. These are shown in Figure 5(f) for profile 1 across the wire and in Figure 5(g) for profile 2 which shows the stray field in vacuum, slightly away from the apex of the wire. For both the electron holography and pixelated DPC measurements the profiles are largely consistent with the simulations. Overall, Figures 5(a-c) reveal a good agreement of both measurements of the projected magnetic induction with the model, in terms of orientation and magnitude of the field. Still, the artifact resulting from diffraction contrast revealed in Figure 2 is visible in the holography measurement in Figure 5(a). However, Figures 5(b) and (f) show that the pixelated DPC fails inside the nanowire. This is due to two main reasons. Firstly, dynamical diffraction contrast shows that the phase approximation assumed for Equation 5 is not valid. Dynamical diffraction contrast is more difficult to remove when using a convergent beam compared to a parallel beam as in holography. In the specific case of this polycrystalline sample, it is not possible to remove them by using the empirical method of finely tuning the specimen tilt until the whole region of interest appears homogeneously bright in the conventional bright-field image. Secondly, the spatial extension of the STEM probe, for which different regions of the probe interact with different regions of the sample with significantly different gradients of potential (see Equation 6), leading to significantly different contributions summed in the diffraction pattern.

For completeness, Figure 5(d) and (e) show the measurements of the projected electric field for holography and pixelated DPC, calculated respectively from the gradient of the electric potential as described by Equations 1 and 3 and from the beam deflection as described by Equation 9. These experimental maps also show a good agreement. In the vacuum, an electric field pointing towards the nanowire is observed, revealing a negative charging of the thin insulating oxide layer at its surface. This field distribution is symmetric relative to the wire axis for holography, for which the specimen quickly reaches a steady state of charge. However, for pixelated DPC, this field distribution is not symmetrical, possibly caused by dynamic charging of the oxide layer when using scanning mode which would induce local modifications of the electric field while the probe is scanning the specimen. At the nanowire location, the strong influence of the MIP is observed, and Figure 5(d) and (e) show basically the opposite of the gradient of the projected thickness of the nanowire (see Equation 6). It highlights the need for a specimen of uniform thickness for measuring electric fields by holography and pixelated DPC.

After this qualitative discussion of the fields, we now compare the sensitivity and spatial resolution of the measurements. In this study, the sensitivity is estimated as the standard deviation of the noise of the measurement. The holography measurements shown in Figure 2 reach a high phase sensitivity of $2\pi/1150$ rad in vacuum (i.e. for the stray fields), associated with a spatial resolution of 3.5 nm. The spatial resolution is determined by the numerical aperture used for the reconstruction of the series of holograms [31]. This sensitivity was made possible by the summation of a large series of electron holograms, which allowed a huge cumulative exposure time of 13 min 20 s. However, performing the numerical differentiation of the magnetic and electric phase images to calculate the experimental holography field maps increases the high-spatial-frequency noise. To limit this effect, Figures 5(a) and (d) were convoluted with a Gaussian kernel that decreases the spatial resolution of the maps to 9 nm. Thus, a high sensitivity of 0.09 T·nm is obtained, instead of 1.80 T·nm when preserving the 3.5 nm spatial resolution of the initial holography reconstruction, see Table 1. Considering the pixelated DPC measurement, the spatial resolution is estimated to be the pixel size of the map which is determined to be 9 nm from the step of the scan, as the electron probe of ∼ 5 nm is smaller. However, the much longer extension of the tail of the probe seems to limit further the resolution of the fields, resulting in an apparent larger nanowire diameter observed in Figures 5(b) and (e). This effect can be explained by the high sensitivity of COM algorithm to the displacement of small fractions of the probe intensity. Nevertheless, the spatial resolution of pixelated DPC is comparable to the holography field maps and is associated with a sensitivity of 24 nrad on the beam deflection in vacuum, which corresponds to a sensitivity of 0.04 T·nm for the magnetic field integrated over the electron beam path. For this pixelated DPC experiment, the sample received a total electron dose of 3.5E5 $e^-$·$nm^{-2}$, to be compared to 1.07E6 $e^-$·$nm^{-2}$ for holography. Processing of only 33 holograms of the series gives a fields sensitivity of 0.15 T·nm and allows to compare the sensitivity of pixelated DPC and holography for the same dose and spatial resolution. These results are summarised in Table 1.

Finally, the higher sensitivity to magnetic induction with pixelated DPC arises from the fact that the measured quantity, the deflection angle of the beam, is related to the fields themselves. It does not involve a derivative as for holography, which is sensitive to the phase. The profiles shown in Figure 5(g) reveal that although pixelated DPC has a higher sensitivity, electron holography has a better accuracy. This is most likely attributed to the artifacts from dynamical charging of the specimen when using the scanning mode and also to the difficulty to calibrate the diffraction patterns very accurately at high camera length. Furthermore, the spatial resolution is better for holography and higher spatial resolutions can be reached at the expense of sensitivity. Of course, the comparison of the two techniques is rather complex as there are always methods of improving sensitivity and spatial resolution. Globally the sensitivity is improved by the use of higher electron dose, via higher beam currents or longer exposure times. However, the important quantity for pixelated DPC is not directly the dose received by the sample but the electron dose collected in each diffraction pattern. The sensitivity of pixelated DPC can be improved by the use of longer camera length and its spatial resolution by higher convergence angle of the probe, but the combination of these two parameters is limited by the physical size of the pixelated detector. Here we present a work that has been performed using an old generation CCD detector. The use of a modern hybrid pixel array detector, designed for pixelated STEM applications, would drastically improve the data collection speed and will allow improved experimental results in other respects.

Pixelated DPC technique is relatively recent, motivated by the emergence of hybrid pixel array detectors and powerful desktop computers associated with fast networks that can manage the large volumes of data that are recorded. However, this technique can be performed in any microscope that is equipped with a camera and suitable software controls to synchronize the STEM scan with the data acquisition, whilst holography requires an electron biprism, which is not always available. Most importantly, a field free reference wave is required for electron holography. In the case of magnetic stray fields, the reference wave is likely to be perturbed by the long-range fields. This can be avoided by using specialized holography microscopes that use multiple biprisms to take the reference far from the region of interest [46]. For the pixelated DPC, the problem of dynamic charging in the specimens seems to be an important problem, which could eventually be improved by the use of sparse acquisition [47]. An additional observation is that for measurements of the fields inside the specimens, it is easier to remove contributions from dynamical diffraction by tilting the specimen when using a plane wave in electron holography, whereas for the use of a convergent beam in pixelated DPC this becomes more difficult. However, these problems can in principle be avoided through the use of precession during pixelated DPC measurements, which has already been demonstrated for the measurement of deformation [48,49] and electric potentials [44,50–52]. Many other improvements can be made for pixelated DPC, such as live data analysis, the use of different algorithms [53] that can measure the shift of the transmitted beam more accurately in the presence of dynamical diffraction within the rigid-shift regime and the live correction of the drift of the specimen for long acquisitions.

**Conclusion**

Pixelated DPC was compared to off-axis electron holography for the mapping of magnetic induction fields at nanometer-scale, using a field-free configuration of the microscope. A method was developed to map quantitatively the magnetic induction and electric field at medium spatial resolution by pixelated DPC. It shows a good quantitative agreement with modelling and electron holography measurements. Whilst holography is sensitive to potentials, pixelated DPC is sensitive to potentials gradients that directly correspond to fields. Pixelated DPC has a higher sensitivity for measurement of the fields but a lower spatial resolution when compared to holography. The main drawback for pixelated DPC is that it is more sensitive to diffraction contrast, due to the convergence angle of the probe, yielding artifacts within thick materials. The main problem for electron holography is that a reference area is required close to the measurement area, making the technique less versatile. These results are encouraging and there are many ways to improve pixelated DPC measurements, as for example precession diffraction to reduce the influence of diffraction contrast or simply the use of a hybrid pixel array detector to improve the sensitivity and acquisition speed.

**Acknowledgements**
D.C. thanks the ERC for funding the Starting Grant no. 306535 "Holoview". These experiments were performed at the NanoCharacterisation PlatForm (PFNC) at Minatec and supported by the "Recherches

Technologiques de Base" Program of the French Ministry of Research. M.S. acknowledges a grant from the Laboratoire d'excellence LANEF in Grenoble (ANR-10-LABX-51-01).

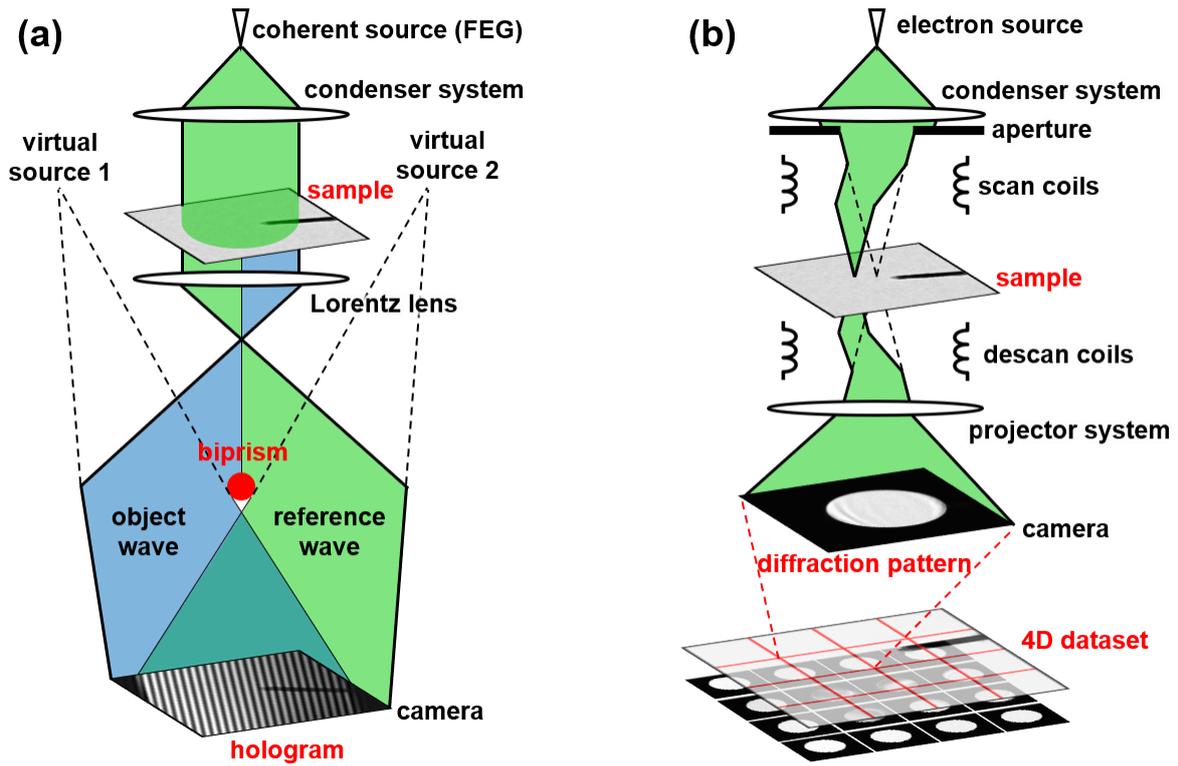

**Figure 1**: Schematics of the experimental setups for (a) off-axis electron holography and for (b) pixelated STEM.

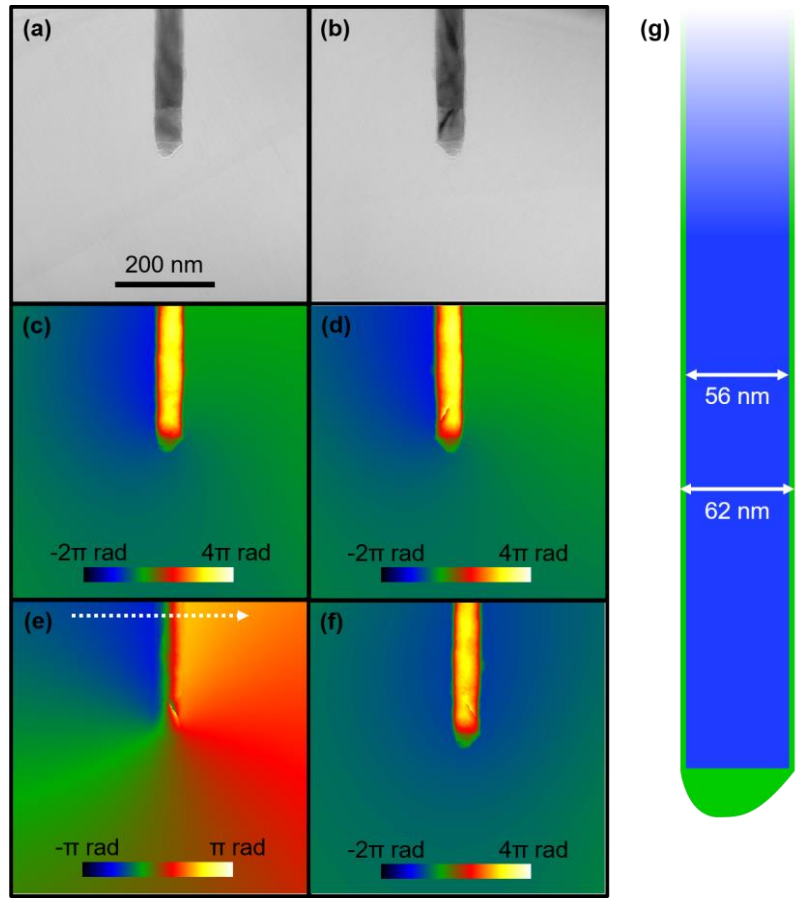

**Figure 2**: Off-axis electron holography measurements. (a) and (b) show amplitude images of the NiFe nanowire obtained before and after flipping the specimen by 180°, respectively. (c) and (d) show the reconstructed phase images of the nanowire before and after flipping the specimen, respectively. (e) shows the magnetic and (f) the electric phase contributions for orientation as in (a), separated using the "flipping method". The profile extracted along the dashed line in (e) is plotted in Figure 3(b). (g) Diagram of the NiFe nanowire (blue) with its oxide shell (green).

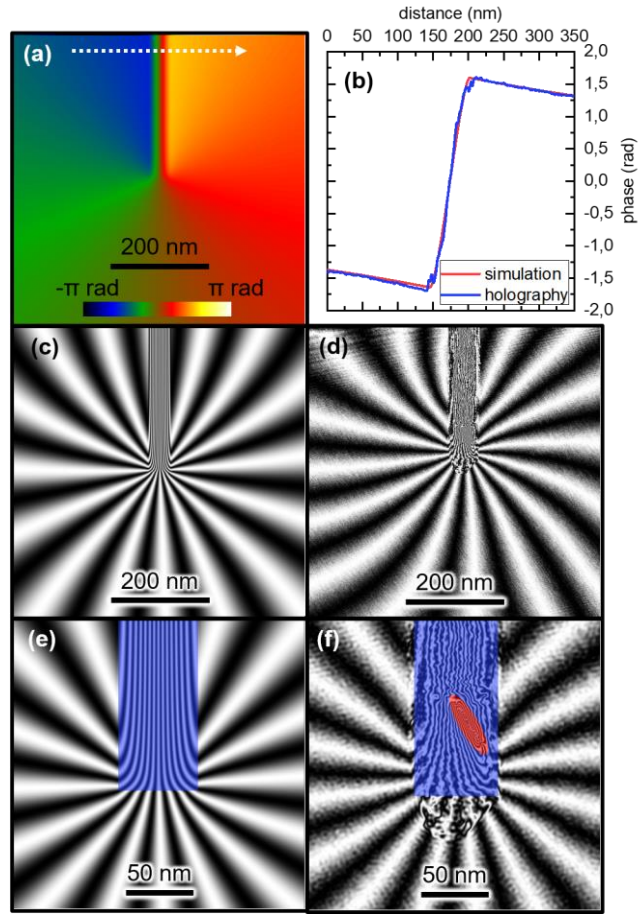

**Figure 3**: (a) Simulation of the magnetic phase shift generated by the NiFe nanowire. (b) Plot of the magnetic phase profiles from the simulation [dashed line in (a)] and measured by holography [dashed line in Figure 2(e)]. Magnetic equipotential lines (c) simulated and (d) measured by holography, displayed as the cosine of 30 times the magnetic phase images, whilst (e) and (f) show the same images enlarged close to the wire apex, respectively. The blue shaded region represents the location of the NiFe nanowire and the red shaded region in (f) reveal an artifact from the measurement.

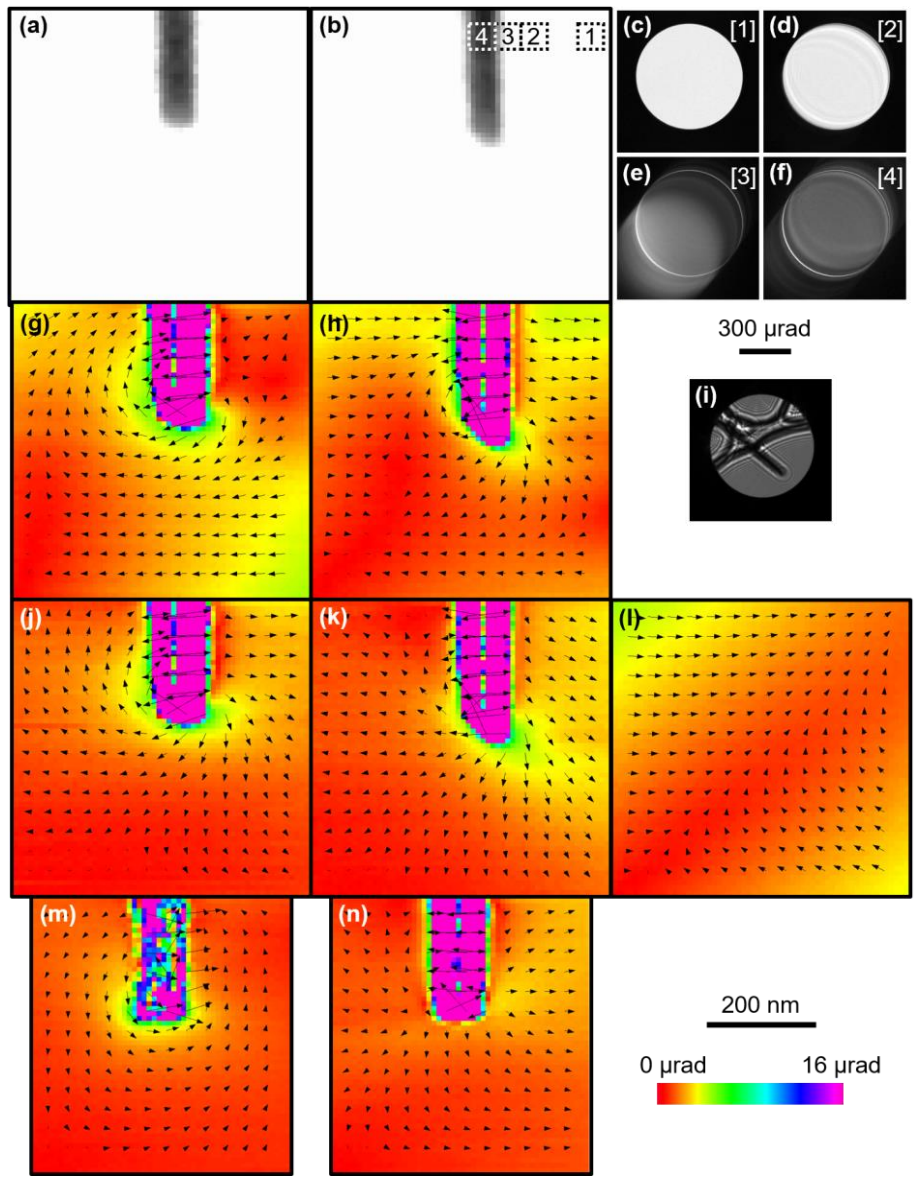

**Figure 4:** Pixelated DPC measurements. (a) and (b) show the virtual bright-field images of the nanowire recorded before and after flipping the specimen by 180°, respectively. (c-f) show the distributions of intensity of the transmitted beam of the diffraction patterns sampled from positions 1, 2, 3 and 4 indicated in (b). (g) and (h) display the maps of the shifts of the beam calculated by the COM technique, respectively before and after flipping the specimen. (i) shows the defocused diffraction image of the nanowire on a lacey carbon grid. (j) and (k) show the beam deflection maps containing both the magnetic and electric contributions of phase gradients, respectively before and after flipping the specimen and after removing the systemic rigid shifts of the beam from the reference scan, displayed in (l). (m) and (n) show the magnetic and electric components of the beam deflection, respectively, separated using the "flipping method". In all maps, the color codes the magnitude of deflection and is restricted to 16 µrad to visualize the details in the vacuum region, while arrows indicate their orientation and magnitude.

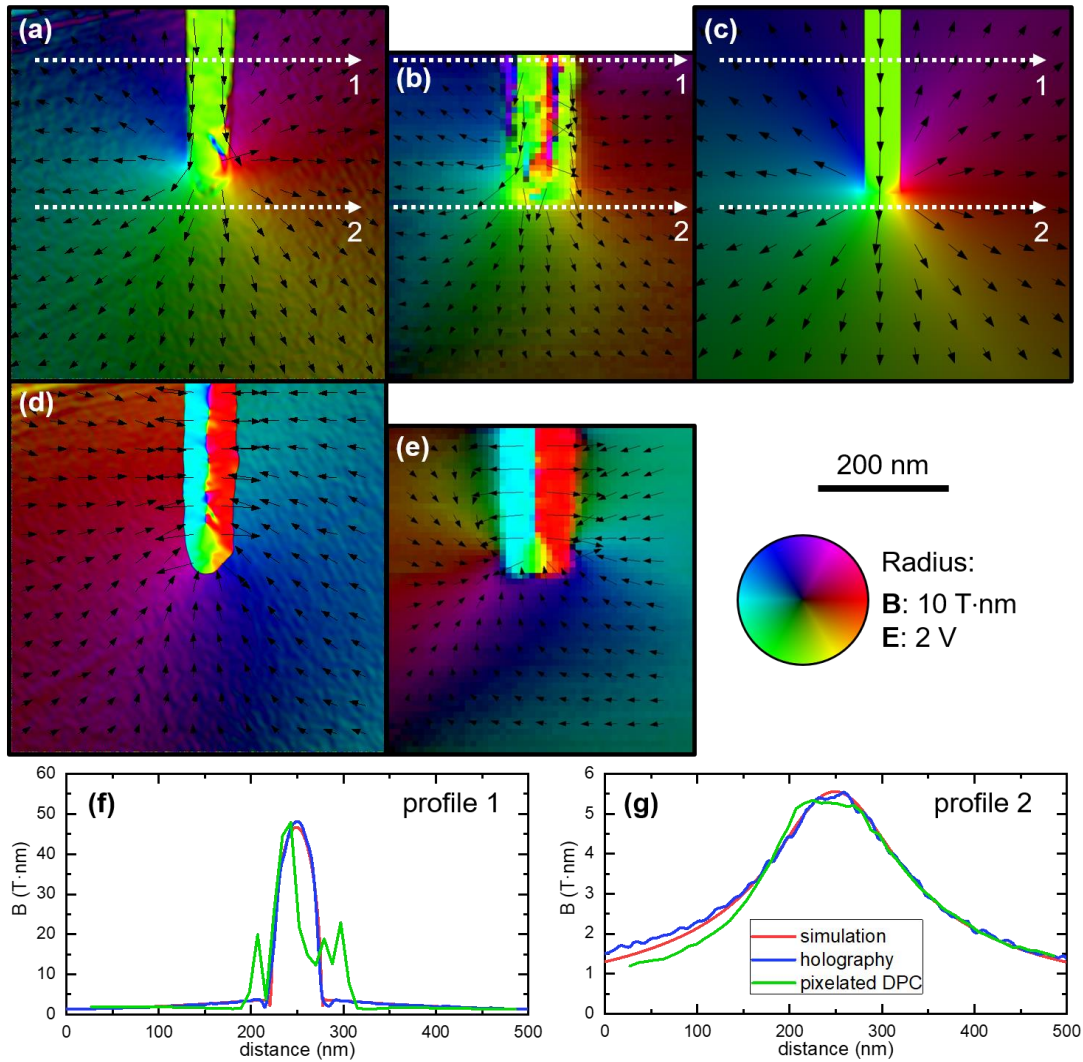

**Figure 5:** Projected magnetic induction maps (a) measured by electron holography, (b) measured by pixelated DPC and (c) simulated. Projected electric field maps (d) measured by electron holography and (e) measured by pixelated DPC. The color wheel encodes the orientation of the fields in its color and the magnitude of the fields in its color saturation. The overprint of arrows further guides the reading. Projections of the magnetic induction field are given in units of T·nm, and projections of electric field in units of V. (f) and (g) show profiles of the magnitude of the integrated magnetic induction extracted along the dashed arrows of (a-c).

**Table 1:** Sensitivity, spatial resolution and electron dose for different field reconstructions of the holography experiment, compared to the pixelated DPC measurement.

|  | Spatial resolution (nm) | Electron dose (e$^-$·nm$^{-2}$) | Sensitivity (T·nm) |
|---|---|---|---|
| Holography (100 holograms) | 3.5 | 1.07E6 | 1.80 |
| Holography (100 holograms & Gaussian filter) | 9 | 1.07E6 | 0.09 |
| Holography (33 holograms & Gaussian filter) | 9 | 3.5E5 | 0.15 |
| Pixelated DPC | ∼ 9 | 3.5E5 | 0.04 |